# The Open Access Advantage Considering Citation, Article Usage and Social Media Attention

Xianwen Wang[*], Chen Liu, Wenli Mao and Zhichao Fang

WISE Lab, Faculty of Humanities and Social Sciences, Dalian University of Technology, Dalian 116085, China

* Corresponding author.

Email address: xianwenwang@dlut.edu.cn; xwang.dlut@gmail.com

Website: http://xianwenwang.com

**Abstract:** In this study, we compare the difference in the impact between open access (OA) and non-open access (non-OA) articles. 1761 *Nature Communications* articles published from 1 Jan. 2012 to 31 Aug. 2013 are selected as our research objects, including 587 OA articles and 1174 non-OA articles. Citation data and daily updated article-level metrics data are harvested directly from the platform of nature.com. Data is analyzed from the static versus temporal-dynamic perspectives. The OA citation advantage is confirmed, and the OA advantage is also applicable when extending the comparing from citation to article views and social media attention. More important, we find that OA papers not only have the great advantage of total downloads, but also have the feature of keeping sustained and steady downloads for a long time. For article downloads, non-OA papers only have a short period of attention, when the advantage of OA papers exists for a much longer time.

## Introduction

Since Lawrence proposed the open access citation advantage (Lawrence 2001), the advantage of open access articles compared to non-open access articles has been debated a lot (Joint 2009; Norris et al. 2008; Davis et al. 2008; Moed 2007). The ways to test the impact advantage of OA not only include comparing the impact factors of OAJ (open access journal) and traditional journals (Antelman 2004) , but also comparing the impact of individual OA articles and non-OA articles appearing in the same non-OA journals (Harnad and Brody 2004). Some studies found that open access leads to obvious citation advantage (Gargouri et al. 2010; Greyson et al. 2009).There are also many other factors affecting citation rates, i.e., papers from different countries published in the same journal may have different citation rates (Akre et al. 2009).

Besides the citation data, there are other novel types of metrics data used and studied by many researchers in recent years. Among of them, article usage metrics and Altmetrics have attracted much attention from bibliometrics scientists (Duy and Vaughan 2006; Davis and Solla 2003; Davis et al. 2008; Kurtz and Bollen 2010; Priem et al. 2010; Piwowar 2013). Three years ago, very few publishers provided usage statistics data for their published articles. This situation has been changed a lot recently. Many publishers and digital libraries begin to provide article-level usage data to public, i.e., ACM Digital Library, Taylor & Francis. Some publishers even go further. For example, PLOS and IEEE Xplore digital library provide article views data for each paper in each month, the nature.com journal platform provides daily page views counts data for each research paper published by *Nature* and other Nature journals, which we call it dynamic usage data (Wang et al. 2014a; Wang et al. 2014b). In our previous study, the main article usage statistics tools provided by publishers are listed (Wang et al. 2014a). There have been more daily updated dynamic usage data sources since our previous study published, for example, in 2014, *Science* and *PNAS* began to provide monthly "Article Usage" data of their published items. With this kind of dynamic article usage data, it's possible for us to trace the realtime research trends when we know what papers are being downloaded by researchers right now (Wang et al. 2013b), to explore the usage patterns of scientific literature with the data of how many times has one specific paper been downloaded each day (Wang et al. 2014a; Wang et al. 2014b). We also examine the time of day when people download articles from Springer. Controlling for the time zone where the request originated, we are able to see how hard scientists work overall (Wang et al. 2013a; Wang et al. 2012).

For the altmetrics data, some academic publishers have integrated Altmetrics data on the article pages in their journals, e.g., Springer, Wiley, Science, Nature, PNAS, etc. Or, we may check the Altmetric score from the website of www.altmetric.com for any article with the digital object identifier (doi), for example, http://www.altmetric.com/details.php?doi=10.1016/j.joi.2012.07.003. With Altmetric data, the societal impact of scientific literature has been studied by some studies (Thelwall et al. 2013; Priem et al. 2012; Kwok 2013; Mounce 2013).

In this study, using citation data, usage data and social media discussion data, our research questions are, how is the temporal evolution of article usage of OA and non-OA articles? How is the difference of the article views between OA and non-OA articles? Could the citation advantage of OA articles be extended to the article views and social media discussion?

**Data and Method**

In this study, we choose *Nature communications* as the test bed, which is a sub-journal of Nature. There are several major reasons for us to choose it. Firstly, *Nature Communications* is an online-only journal, which could totally exclude the effects of the article views of hard copy edition of journal. Secondly, unlike other non-open access journals that have only a few OA articles, before it became fully open access in September, 2014, *Nature Communications* not only had non-OA papers, but also had a large amount of OA papers. With the comparable data of OA and non-OA articles from the same journal, we could make a better comparative analysis between OA and non-OA articles. Thirdly, like all Nature journals, *Nature Communications* provide article metrics data for each research article. As Figure 1 shows, the metrics data includes citation data, online attention data and article page views data. More important, the page views data is not only restricted to the total article views of an article, but also daily updated.

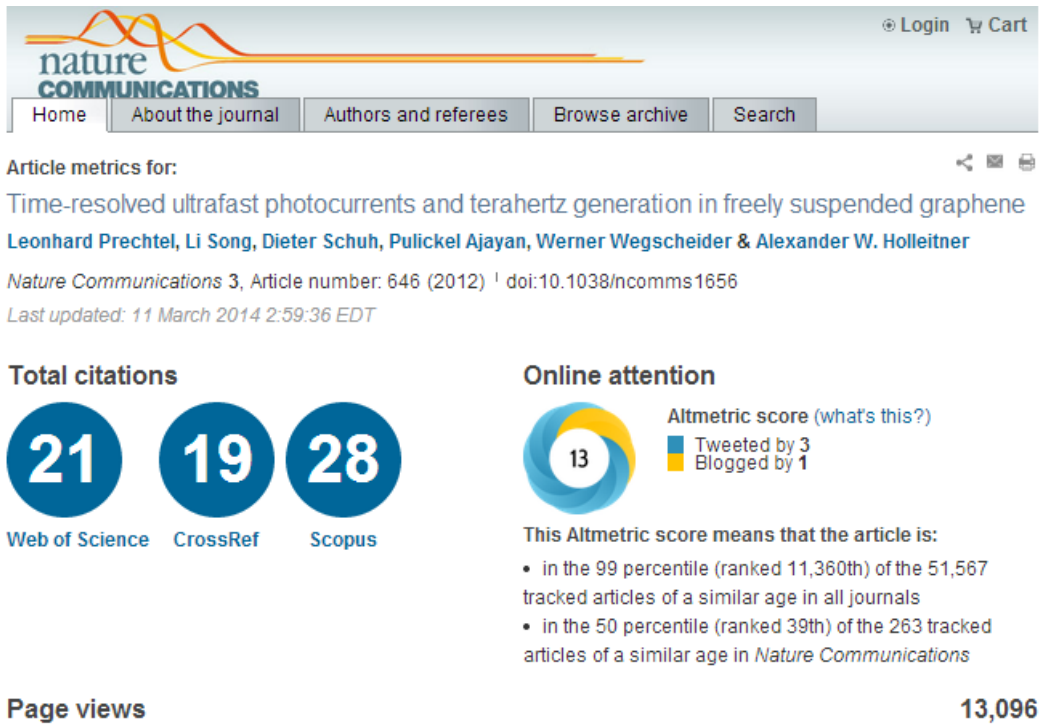

Figure 1. Article metrics of one *Nature Communications* publications.

To ensure all the articles have enough time to accumulate the metrics data, we choose the articles which were published between 1 Jan. 2012 and 31 Aug. 2013. Therefore, we get 1761 articles in total, including 587 OA articles and 1174 non-OA articles; the number of OA articles is approximate half the non-OA articles. For each article, the daily updated metrics data is collected and parsed into a SQL database designed for our purpose. All the data is processed and analyzed in the SQL database. Articles published in January 2012 have the longest publication history over 700 days in our dataset, when articles published in August 2013 have the shortest publication history, about 6 months.

Figure 2 shows the accumulated page views for each article. The blue lines indicate the OA papers, when the black line is the average of all OA papers (blue lines); and the orange lines indicate the non-OA papers, when the red line is the average of all non-OA papers

(orange lines). As shown in Figure 2, most blue lines (OA papers) are higher than the orange lines (non-OA paper), which is also reflected by the black line (average OA paper) and red line (average non-OA paper).

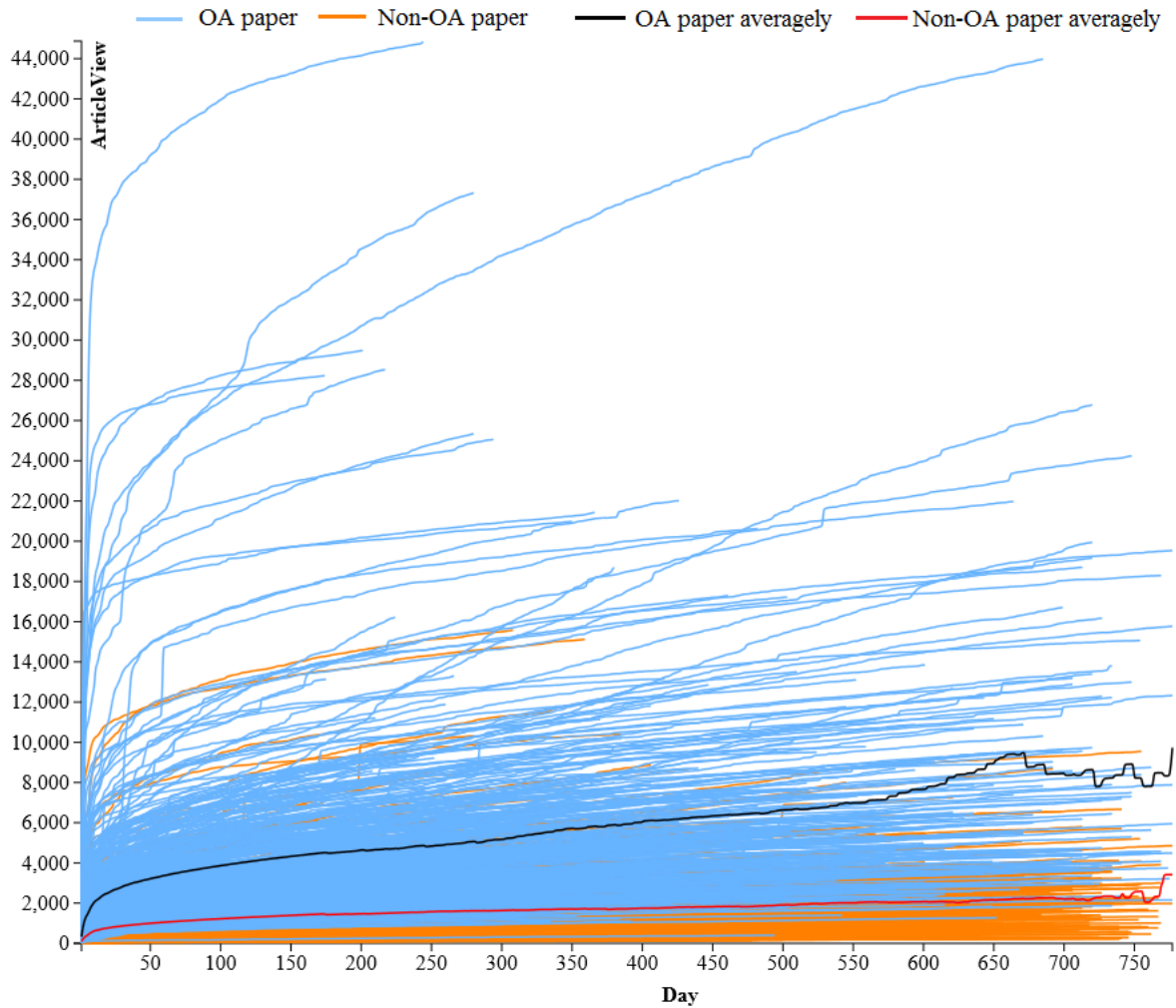

**Figure 2. Accumulated page views of *Nature Communication* articles**

**Results**

*Static comparison*
We choose three indicators to compare the OA and non-OA articles, which are citation, article views and social media discussion (twitter and facebook). 1761 articles are classified into five groups according to the publish date with a 4-month interval, for example, articles published from January to April, 2012 are in the same group, articles published from May to August, 2012 are in the next group and so on, as Table 1 shows.

Table 1 Comparison between OA and non-OA articles

|  | Article views | | | Citation | | | Social media | | |
|---|---|---|---|---|---|---|---|---|---|
|  | OA | Non-OA | OA/Non-OA | OA | Non-OA | OA/Non-OA | OA | Non-OA | OA/Non-OA |
| Jan-Apr,2012 | 10073.27 | 2291.19 | 4.40 | 18.37 | 12.1 | 1.52 | 2.91 | 2.38 | 1.22 |
| May-Aug,2012 | 5919.56 | 2088.59 | 2.83 | 13.04 | 9.34 | 1.40 | 2.82 | 2.01 | 1.40 |
| Sep-Dec,2012 | 6419.49 | 1755.83 | 3.66 | 6.38 | 4.72 | 1.35 | 3.41 | 2.30 | 1.48 |
| Jan-Apr,2013 | 4876.05 | 1907.27 | 2.56 | 3.4 | 2.74 | 1.24 | 2.81 | 2.22 | 1.27 |
| May-Aug,2013 | 5408.87 | 1909.62 | 2.83 | 1.09 | 0.98 | 1.11 | 3.15 | 2.14 | 1.47 |

Figure 3 shows the comparison results of page views between OA and non-OA articles published in the same period. An obvious phenomenon is that the average page views of non-OA articles in all five groups show a similar result, vacillating in a range of 1750 to 2300, when the range of average page views for OA articles is 4800 to 6500, which is 2.5 to 3.6 times of the corresponding non-OA group. For the "Jan-Apr, 2012" group, articles in which have been published for about two years, the gap is considerable. OA articles in this group have 10073.27 page views on average, about 4.4 times of the corresponding non-OA group.

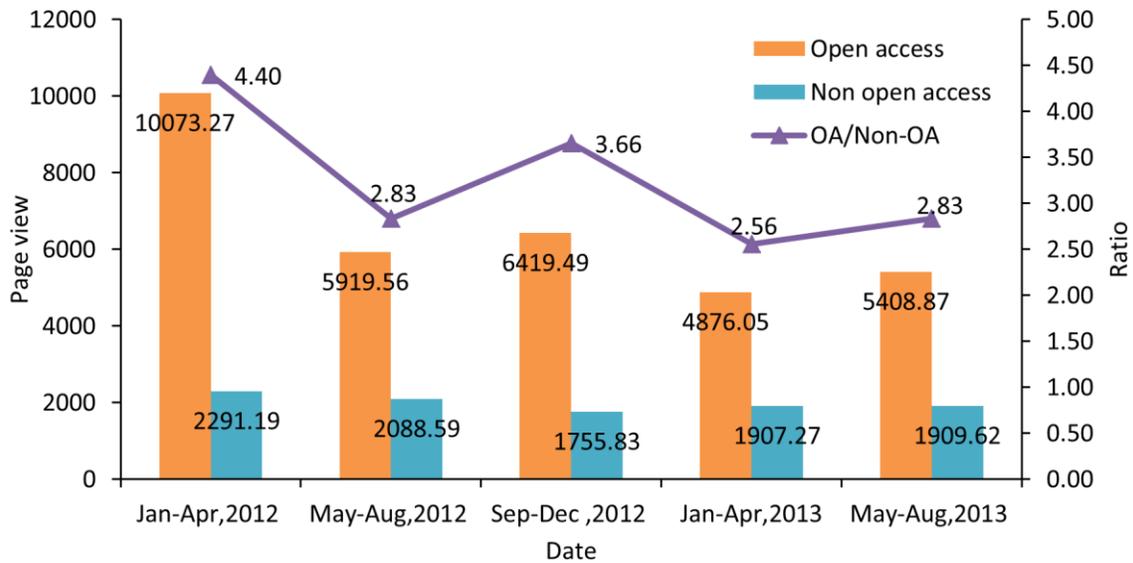

**Figure 3. Comparison of average page views between OA and non-OA articles.**

453 of all the 587 OA articles have at least one citation, and the citedness calculated by the number of OA articles with at least one citation divided by the number of all OA articles is 77.17%. Meanwhile, for the 1774 non-OA articles, 808 articles are cited at least one time, when the citedness is only 68.82%. OA articles have a clear citation advantage which confirms the hypothesis of Garfield (Garfield 2004). For the comparison of citation counts, the difference is not as considerable as the comparison result of page views. As Figure 4 shows, for the last columns, the "May-Aug, 2013" group, articles in this group have the shortest publication history of only 7 to 11 months, OA articles have been cited 1.09 times on average, when the average citations for non-OA articles is 0.98, the results of both OA and non-OA articles are quite close. However, for the "Jan-Apr, 2013" group on the left side of the "May-Aug, 2013" group, when the former has a longer publication history than the later group, the ratio of OA to non-OA articles rise to 1.24. If we examine the result from right to left, as the publication history gets longer, the ratio increases, which means the OA citation advantage becomes more and more apparent. For the most left columns, OA articles have 18.37 citations on average, which is about 1.52 times of non-OA articles.

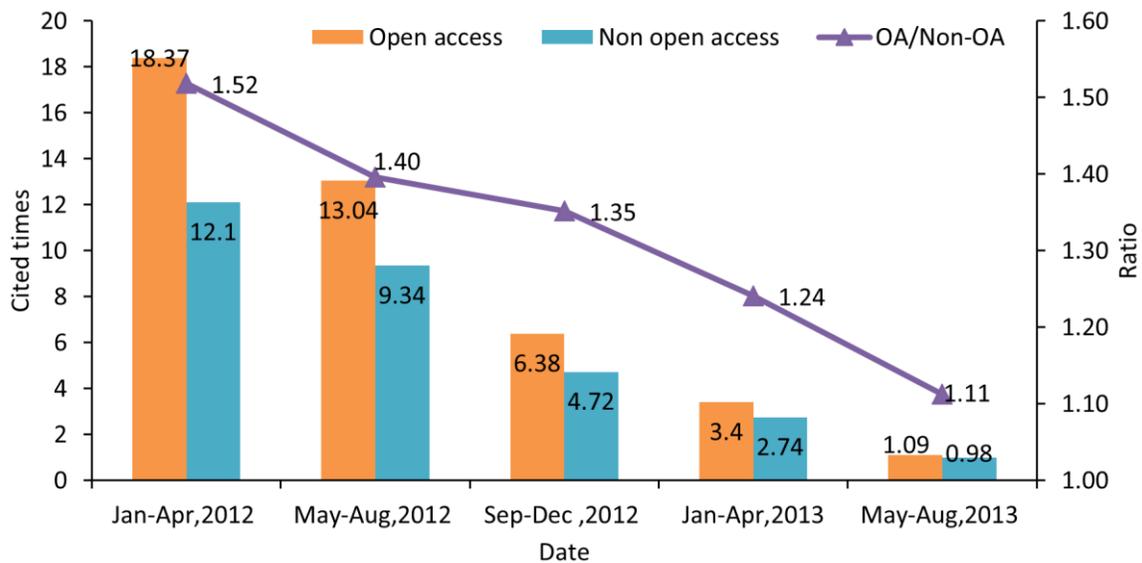

**Figure 4. Comparison of citation between OA and non-OA articles.**

As shown in Figure 5, for all the groups, the average number of twitter and facebook of OA articles is 2.8 to 3.4, which is slightly more than the number of non-OA articles, when the ratio of OA articles to non-OA articles is 1.27 to 1.48. OA articles attract a litter more social media attention than non-OA articles.

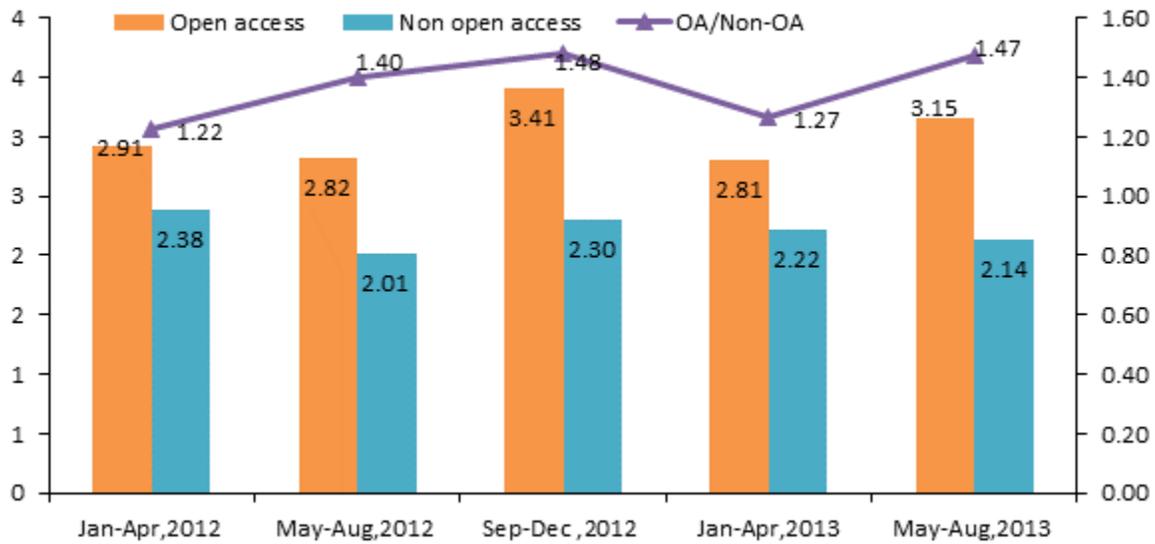

**Figure 5. Comparison of social media attention between OA and non-OA articles.**

*Dynamic comparison*

In order to examine the dynamic changes of the influence of OA and non-OA articles, daily page view data of each article is collected and analyzed here. Figure 6 shows the average accumulation page views for OA and non-OA articles. The x axis indicates the published days, when the y axis represents the accumulated page views. If one paper is published on January 1, 2012, we consider the day as day 0, and January 2, 2012 as day 1 and so on. As Figure 6 shows, the most left part of both two curves is rather smooth, when the tail end of the curves fluctuates noticeably. The main reason is that few articles have been published for over 600 days, the average result would be more fluctuated with less samples. However, it seems pretty clear that the gap between the OA curve and non-OA curve widens as the publication history becomes longer.

Here we design another strategy to compare the difference of dynamic evolution of daily downloads between OA and non-OA articles. Both the two curves are broken into three segments. Data from day 0 to 30 is grouped into the left piece, OA(0-30). Data from day 31 to 600 is grouped into the middle piece, OA(31-600), when the rest data belongs to the right piece, OA(601-776).

In the first period of day 0 to 30, during the first month when a new issue is available, newly published papers attract researchers' most attention, so it is not difficult to understand why the accumulated curves rise so quickly during the first period, as the OA(0-

30) piece shows. During the second period of day 31 to 600, the middle piece of both the two curves shows steady rise, however, there is huge difference between the two middle pieces. For the non-OA curve, the middle piece is rather flat, when the corresponding piece of the OA curve slopes up noticeably. To better compare the difference of the two curves, two liner trend line are superimposed to the middle pieces, when the upward steepness, represented by the slope of the trend line, describes the degree of increase. As Figure 6 shows, the slope of the trend line of the OA curve is 7.87, when the slope of the non-OA curve is only 1.74. The steeper trend line with greater slope indicates the sustained and steady growth of accumulated page views of OA articles. In contrast, for the non-OA articles, after a rapid growth of the first 30-days period, there are few new views during the following long period.

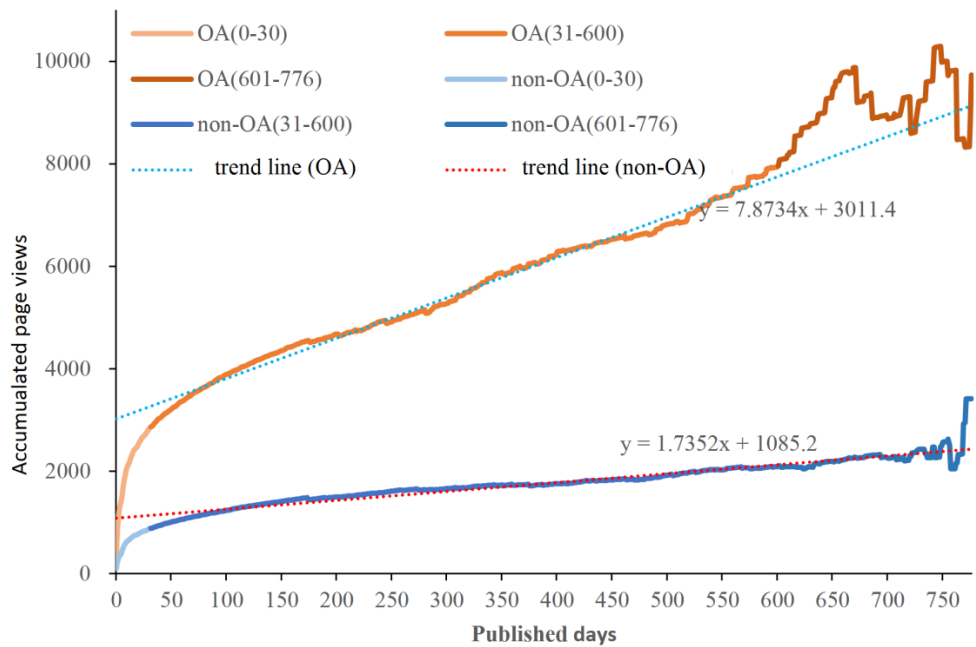

**Figure 6. Comparison of accumulation page view between OA and non-OA articles.**

**Conclusion and discussion**

In this study, using the article usage data, citation data and altmetrics data of *Nature Communication* publications, we compare the difference of OA and non-OA articles. From the perspective of static comparison, we confirm the hypothesis of OA citation advantage, when the OA advantage could also be extended to the usage metrics. OA articles get much more attention than non-OA papers. More importantly, from the perspective of temporal-dynamic comparison of the daily usage metrics data, we find that the accumulated advantage of OA articles increases greatly as the publish time get longer. OA articles could attract sustained and steady attention even a long time after publish. In contrast, for the non-OA articles, most attention only occurs in the first 30-day period (one month), after that, the new downloads become rare. In summary, the OA advantage exists not only for citation, but also for article usage. Compared with the short period attention for non-OA papers, the OA advantage of article usage exists for a much longer time period.

**Limitation**

According our previous studies, downloading rates for OA and non-OA articles are very dynamic, e.g., OA articles are able to attract immediate views in a short time period, while non-OA articles decrease downloading numbers much faster and more dramatically (Wang, et al., 2014). In this study, the result is incapable of reflecting the dynamics, which may be the result of the relatively small size of research samples.

It has been recognized in the OA papers that citation measures are reliable for suggesting the influence and impact of open access status, though they are very slow to collect. Besides the data format of citation counts, in this study, we pay more attention to altmetrics. Compared with the long time needed to accumulate enough citations to make evaluations, altmetrics data, including download statistics, are rapid to collect. Altmetrics data is very useful to make fast evaluation to newly published papers, however, only using altmetrics data to evaluate articles may be misleading. Many variables need to be considered. For example, did the authors of the OA articles intentionally select their best articles for free access? Will authors' seniority has a say on article-level views and citations? Will authors' origin of country plays a role?

In this study, the page views data reflect only the article usage on the nature.com journal platform, not including article usage data from other third-party services. This limitation may affect the result. For example, authors may distribute their non-OA papers through their personal websites or some self-archived platforms, e.g, arXiv. These self-archiving papers also could be crawled by Google Scholar, and may promote the article usage even the published papers are non-OA. However, these data are very difficult to find out and collect.

**Acknowledgments**

This work was supported by the project of ''National Natural Science Foundation of China'' (61301227), and the project of "Growth Plan of Distinguished Young Scholar in Liaoning Province"(WJQ2014009). We really appreciate the suggestions from the anonymous reviewers.

# References

Akre, O., Barone-Adesi, F., Pettersson, A., Pearce, N., Merletti, F., & Richiardi, L. (2009). Differences in citation rates by country of origin for papers published in top-ranked medical journals: do they reflect inequalities in access to publication? *Journal of epidemiology and community health, 65*(2), 119-123.

Antelman, K. (2004). Do open-access articles have a greater research impact? *College & research libraries, 65*(5), 372-382.

Davis, P. M., Lewenstein, B. V., Simon, D. H., Booth, J. G., Connolly, M. J., & Godlee (2008). Open access publishing, article downloads, and citations: randomised controlled trial. *BMJ: British Medical Journal, 337*(7665), 343–345.

Davis, P. M., & Solla, L. R. (2003). An IP-level analysis of usage statistics for electronic journals in chemistry: Making inferences about user behavior. *Journal of the American Society for Information Science and Technology, 54*(11), 1062-1068.

Duy, J., & Vaughan, L. (2006). Can electronic journal usage data replace citation data as a measure of journal use? An empirical examination. *Journal of Academic Librarianship, 32*(5), 512-517, doi:10.1016/j.acalib.2006.05.005.

Garfield, E. (2004). What is the threshold for open access Nirvana? http://users.ecs.soton.ac.uk/harnad/Hypermail/Amsci/3427.html2.

Gargouri, Y., Hajjem, C., Larivière, V., Gingras, Y., Carr, L., Brody, T., et al. (2010). Self-selected or mandated, open access increases citation impact for higher quality research. *Plos One, 5*(10), e13636.

Greyson, D., Morgan, S., Hanley, G., & Wahyuni, D. (2009). Open access archiving and article citations within health services and policy research. *Journal of the Canadian Health Libraries Association, 30*(2), 51-58.

Harnad, S., & Brody, T. (2004). Comparing the impact of open access (OA) vs. non-OA articles in the same journals. *D-lib Magazine, 10*(6), http://www.dlib.org/dlib/june04/harnad/06harnad.html. Accessed February 16, 2015.

Joint, N. (2009). The Antaeus column: does the "open access" advantage exist? A librarian's perspective. *Library review, 58*(7), 477-481.

Kurtz, M. J., & Bollen, J. (2010). Usage bibliometrics. *Annual review of information science and technology, 44*(1), 1-64.

Kwok, R. (2013). Research impact: Altmetrics make their mark. *Nature, 500*(7463), 491-493.

Lawrence, S. (2001). Online or invisible. *Nature, 411*(6837), 521.

Moed, H. F. (2007). The effect of "open access" on citation impact: An analysis of ArXiv's condensed matter section. *Journal of the American Society for Information Science and Technology, 58*(13), 2047-2054.

Mounce, R. (2013). Open access and altmetrics: Distinct but complementary. *Bulletin of the American Society for Information Science and Technology, 39*(4), 14-17.

Norris, M., Oppenheim, C., & Rowland, F. (2008). The citation advantage of open-access articles. *Journal of the American Society for Information Science and Technology, 59*(12), 1963-1972.

Piwowar, H. (2013). Altmetrics: Value all research products. *Nature, 493*(7431), 159-159.

Priem, J., Piwowar, H. A., & Hemminger, B. M. (2012). Altmetrics in the wild: Using social media to explore scholarly impact. *arXiv preprint arXiv:1203.4745*.

Priem, J., Taraborelli, D., Groth, P., & Neylon, C. (2010). Altmetrics: A manifesto. http://altmetrics.org/manifesto/. Accessed August 4 2014.

Thelwall, M., Haustein, S., Larivière, V., & Sugimoto, C. R. (2013). Do altmetrics work? Twitter and ten other social web services. *Plos One, 8*(5), e64841.

Wang, X. W., Mao, W. L., Xu, S. M., & Zhang, C. B. (2014a). Usage history of scientific literature: Nature metrics and metrics of Nature publications. *Scientometrics, 98*(3), 1923-1933, doi:10.1007/s11192-013-1167-5.

Wang, X. W., Peng, L., Zhang, C. B., Xu, S. M., Wang, Z., Wang, C. L., et al. (2013a). Exploring scientists' working timetable: A global survey. *Journal of Informetrics, 7*(3), 665-675, doi:10.1016/j.joi.2013.04.003.

Wang, X. W., Wang, Z., Mao, W. L., & Liu, C. (2014b). How far does scientific community look back? *Journal of Informetrics, 8*(3), 562-568, doi:http://dx.doi.org/10.1016/j.joi.2014.04.009.

Wang, X. W., Wang, Z., & Xu, S. M. (2013b). Tracing scientist's research trends realtimely. *Scientometrics, 95*(2), 717-729, doi:10.1007/s11192-012-0884-5.

Wang, X. W., Xu, S. M., Peng, L., Wang, Z., Wang, C. L., Zhang, C. B., et al. (2012). Exploring scientists' working timetable: Do scientists often work overtime? *Journal of Informetrics, 6*(4), 655-660, doi:10.1016/j.joi.2012.07.003.